\documentclass[5p]{elsarticle}
\usepackage[utf8]{inputenc}
\usepackage{graphicx}
\usepackage{multirow}
\usepackage{amssymb}
\usepackage{mathtools}
\usepackage{listings}
\usepackage[caption=false]{subfig}
\usepackage{balance}
\usepackage{epstopdf}
\usepackage[ 
  bookmarks = true,
  bookmarksopen = true,
  bookmarksnumbered = true,
  breaklinks = true,
  linktocpage,
  pagebackref = false,
  colorlinks = true,
  linkcolor = blue,
  urlcolor  = blue,
  citecolor = red,
  anchorcolor = green,
  hyperindex = true,
  hyperfigures
  ]{hyperref}
\usepackage{url}

\newcommand{\csubfloat}[2][]{%
  \makebox[0pt]{\subfloat[#1]{#2}}}
\newcommand{\e}[1]{{\mathbb E}\left[ #1 \right]}

\newcommand{\superscript}[1]{\ensuremath{^{\textrm{#1}}}}
\def\uc3m{\superscript{*}}

\journal{Computer Communications}

\begin{document}

\begin{frontmatter}

\title{On the energy efficiency of rate and transmission power control in 802.11}

\author[uc3m]{Iñaki~Ucar\corref{corrauth}}
\cortext[corrauth]{Corresponding author}
\ead{inaki.ucar@uc3m.es}

\author[imdea]{Carlos~Donato}
\ead{carlos.donato@imdea.org}

\author[uc3m]{Pablo~Serrano}
\ead{pablo@it.uc3m.es}

\author[nec]{Andres~Garcia-Saavedra}
\ead{andres.garcia.saavedra@neclab.eu}

\author[uc3m,imdea]{Arturo~Azcorra}
\ead{azcorra@it.uc3m.es}

\author[uc3m,imdea]{Albert~Banchs}
\ead{banchs@it.uc3m.es}

\address[uc3m]{Universidad Carlos III de Madrid, 28911 Leganés, Spain}
\address[imdea]{IMDEA Networks Institute, 28918 Leganés, Spain}
\address[nec]{NEC Labs Europe, 69115 Heidelberg, Germany}

\tnotetext[copy]{\textcopyright 2017. This manuscript version is made available under the \href{http://creativecommons.org/licenses/by-nc-nd/4.0/}{CC-BY-NC-ND 4.0 license}. DOI: \href{http://doi.org/10.1016/j.comcom.2017.07.002}{10.1016/j.comcom.2017.07.002}}


\begin{abstract}

Rate adaptation and transmission power control in 802.11 WLANs have received a lot of attention from the research community, with most of the proposals aiming at maximising throughput based on network conditions. Considering energy consumption, an implicit assumption is that optimality in throughput implies optimality in energy efficiency, but this assumption has been recently put into question. In this paper, we address via analysis, simulation and experimentation the relation between throughput performance and energy efficiency in multi-rate 802.11 scenarios. We demonstrate the trade-off between these performance figures, confirming that they may not be simultaneously optimised, and analyse their sensitivity towards the energy consumption parameters of the device. We analyse this trade-off in existing rate adaptation with transmission power control algorithms, and discuss how to design novel schemes taking energy consumption into account.

\end{abstract}


\end{frontmatter}

\section{Introduction}
\label{sec:introduction}

In recent years, along with the growth in mobile data applications and the corresponding traffic volume demand, we have witnessed an increased attention towards ``green operation'' of  networks, which is required to support a sustainable growth of the communication infrastructures. For the case of wireless communications, there is the added motivation of a limited energy supply (i.e., batteries), which has triggered a relatively large amount of work on energy efficiency \cite{survey}. It turns out, though, that energy efficiency and performance do not necessarily come hand in hand, as some previous research has pointed out \cite{tradeoff, balancing}, and that a criterion may be required to set a proper balance between them.

This paper is devoted to the problem of rate adaptation (RA) and transmission power control (TPC) in 802.11 WLANs from the energy consumption's perspective. RA algorithms are responsible for selecting the most appropriate modulation and coding scheme (MCS) to use, given an estimation of the link conditions, and have received a vast amount of attention from the research community (see e.g. \cite{biaz2008, h-rca} and references therein). In general, the challenge lies in distinguishing between those loses due to collisions and those due to poor radio conditions, because they should trigger different reactions. In addition, the performance figure to optimise is commonly the throughput or a related one such as, e.g., the time required to deliver a frame.

On the other hand, network densification is becoming a common tool to provide better coverage and capacity. However, densification brings new problems, especially for 802.11, given the limited amount of orthogonal channels available, which leads to performance and reliability issues due to RF interference. In consequence, some RA schemes also incorporate TPC, which tries to minimise the transmission power (TXP) with the purpose of reducing interference between nearby networks. As in the case of ``vanilla'' RA, the main performance figure to optimise is also throughput.

It is generally assumed that optimality in terms of throughput also implies optimality in terms of energy efficiency. However, some previous work \cite{Li2012, khan2013} has shown that throughput maximisation does not result in energy efficiency maximisation, at least for 802.11n. However, we still lack a proper understanding of the causes behind this ``non-duality'', as it may be caused by the specific design of the algorithms studied, the extra consumption caused by the complexity of MIMO techniques, or any other reason. In fact, it could be an inherent trade-off given by the power consumption characteristics of 802.11 interfaces, and, if so, RA-TPC techniques should not be agnostic to this case.

This work tackles the latter question from a formal standpoint. A question which, to the best of the authors' knowledge, has never been addressed in the literature. For this purpose, and with the aim of isolating the variables of interest, we present a joint goodput (i.e., the throughtput delivered on top of 802.11) and energy consumption model for single 802.11 spatial streams in the absence of interfering traffic. Packet losses occur due to poor channel conditions and RA-TPC can tune only two variables: MCS and TXP.

Building on this model, we provide the following contributions: ($i$)~we demonstrate through an extensive numerical evaluation that energy consumption and throughput performance are different optimisation objectives in 802.11, and not only an effect of MIMO or certain algorithms' suboptimalities; ($ii$)~we analyse the relative impact of each energy consumption component on the resulting performance of RA-TPC, which serves to identify the critical factors to consider for the design of RA-TPC algorithms; ($iii$)~we experimentally validate our numerical results; and ($iv$)~we assess the performance of several representative RA-TPC algorithms from the energy consumption's perspective.

The rest of this paper is organised as follows. In Section~\ref{sec:models}, we develop the theoretical framework: a joint goodput-energy model built around separate previous models. In Section~\ref{sec:results}, we provide a detailed analysis of the trade-off between energy efficiency and maximum goodput, including a discussion of the role of the different energy parameters involved. We support our numerical analysis with experimental results in Section~\ref{sec:experiments}. Section~\ref{sec:simulations} explores the performance of RA-TPC algorithms from the energy consumption's perspective. Finally, Section~\ref{sec:conclusions} summarises the paper.

\section{Joint goodput-energy model}\label{sec:models}

In this section, we develop a joint goodput-energy model for a single 802.11 spatial stream and the absence of interfering traffic. It is based on previous studies about goodput and energy consumption of wireless devices. As stated in the introduction, the aim of this model is the isolation of the relevant variables (MCS and TXP) to let us delve in the relationship between goodput and energy consumption optimality in the absence of other effects such as collisions or MIMO.

Beyond this primary intent, it is worth noting that these assumptions conform with real-world scenarios in the scope of recent trends in the IEEE 802.11 standard development, namely, the amendments 11ac and 11ad, where device-to-device communications (mainly through beamforming and MU-MIMO) are of paramount importance.

\subsection{Goodput model}

We base our study on the work by Qiao \textit{et al.} \cite{Qiao2002}, which develops a robust goodput model that meets the established requirements. This model analyses the IEEE 802.11a Distributed Coordination Function (DCF) over the assumption of an AWGN (Additive White Gaussian Noise) channel without interfering traffic.

Let us briefly introduce the reader to the main concepts, essential to our analysis, of the goodput model by Qiao \textit{et al.}. Given a packet of length $l$ ready to be sent, a frame retry limit $n_\mathrm{max}$ and a set of channel conditions $\hat{s}=\{s_1, \ldots, s_{n_\mathrm{max}}\}$ and modulations $\hat{m}=\{m_1, \ldots, m_{n_\mathrm{max}}\}$ used during the potential transmission attempts, the \emph{expected effective goodput} $\mathcal{G}$ is modelled as the ratio between the expected delivered data payload and the expected transmission time as follows:
\begin{align}\label{goodput}
 \mathcal{G}(l, \hat{s}, \hat{m}) = \frac{\e{\mathrm{data}}}{\e{\mathcal{D}_\mathrm{data}}} = \frac{\Pr[\mathrm{succ} \mid l, \hat{s}, \hat{m}]\cdot l}{\e{\mathcal{D}_\mathrm{data}}}
\end{align}

\noindent where $\Pr[\mathrm{succ} \mid l, \hat{s}, \hat{m}]$ is the probability of successful transmission conditioned to $l, \hat{s}, \hat{m}$, given by Eq.~(5) in \cite{Qiao2002}. This model is valid as long as the coherence time is equal or greater than a single retry, i.e., the channel condition $s_i$ is constant.

The expected transmission time is defined as follows:
\begin{align}\label{Ddata}
 \e{\mathcal{D}_\mathrm{data}} = \left(1 - \Pr[\mathrm{succ} \mid l, \hat{s}, \hat{m}]\right) \cdot \mathcal{D}_{\mathrm{fail} \mid l, \hat{s}, \hat{m}} \\
 + \Pr[\mathrm{succ} \mid l, \hat{s}, \hat{m}] \cdot \mathcal{D}_{\mathrm{succ} \mid l, \hat{s}, \hat{m}} \nonumber
\end{align}

\noindent where
\begin{align}\label{Dsucc}
 \mathcal{D}_{\mathrm{succ} \mid l, \hat{s}, \hat{m}} = &\sum_{n=1}^{n_\mathrm{max}} \Pr[n \mathrm{~succ} \mid l, \hat{s}, \hat{m}] \cdot \biggl\lbrace \sum_{i=2}^{n_\mathrm{max}} \left[\overline{T}_\mathrm{bkoff}(i)\right.\biggr. \nonumber\\
 &+ \left.T_\mathrm{data}(l, m_i) + \overline{\mathcal{D}}_\mathrm{wait}(i)\right] \nonumber\\
 &+ \overline{T}_\mathrm{bkoff}(1) + T_\mathrm{data}(l, m_1) + T_\mathrm{SIFS} \nonumber\\
 &+ \biggl.T_\mathrm{ACK}(m'_n) + T_\mathrm{DIFS} \biggr\rbrace
\end{align}

\noindent is the average duration of a successful transmission and
\begin{align}\label{Dfail}
 \mathcal{D}_{\mathrm{fail} \mid l, \hat{s}, \hat{m}} = &\sum_{i=1}^{n_\mathrm{max}} \left[\overline{T}_\mathrm{bkoff}(i)\right. \\
 &+ \left.T_\mathrm{data}(l, m_i) + \overline{\mathcal{D}}_\mathrm{wait}(i+1)\right] \nonumber
\end{align}

\noindent is the average time wasted during the $n_\mathrm{max}$ attempts when the transmission fails.

$\Pr[n \mathrm{~succ} \mid l, \hat{s}, \hat{m}]$ is the probability of successful transmission at the $n$-th attempt conditioned to $l, \hat{s}, \hat{m}$, and $\overline{\mathcal{D}}_\mathrm{wait}(i)$ is the average waiting time before the $i$-th attempt. Their expressions are given by Equations~(7) and (8) in \cite{Qiao2002}. The transmission time ($T_\mathrm{data}$), ACK time ($T_\mathrm{ACK}$) and average backoff time ($\overline{T}_\mathrm{bkoff}$) are given by Eq.~(1)--(3) in \cite{Qiao2002}. Finally, $T_\mathrm{SIFS}$ and $T_\mathrm{DIFS}$ are 802.11a parameters, and they can be found also in Table~2 in \cite{Qiao2002}.

\subsection{Energy consumption model}

The selected energy model is our previous work of \cite{Serrano2014}, which has been further validated via ad-hoc circuitry and specialised hardware \cite{deseeding} and, to the best of our knowledge, stands as the most accurate energy model for 802.11 devices published so far, because it accounts not only the energy consumed by the wireless card, but the consumption of the whole device. While classical models focused on the wireless interface solely, this one demonstrates empirically that the energy consumed by the device itself cannot be neglected as a device-dependent constant contribution. Conversely, devices incur an energy cost derived from the frame processing, which may impact the relationship that we want to evaluate in this paper.

The energy model is a multilinear model articulated into three main components:
\begin{align}\label{eq:power}
 \overline{P}(\tau_i, \lambda_i) = \underbracket{\rho_\mathrm{id} + \sum_{i\in\mathrm{\{tx,rx\}}} \rho_i \tau_i}_{\text{classical model}} + \sum_{i\in\mathrm{\{g,r\}}} \gamma_{\mathrm{x}i} \lambda_i
\end{align}

\noindent where the first two addends correspond to the classical model and the third is the contribution described in \cite{Serrano2014}. These components are the following:
\begin{itemize}
 \item A platform-specific baseline power consumption that accounts for the energy consumed just by the fact of being powered on, but with no network activity. This component is commonly referred to as \emph{idle consumption}, $\rho_\mathrm{id}$.
 \item A component that accounts for the energy consumed in transmission, which linearly grows with the airtime percentage $\tau_\mathrm{tx}$, i.e., $\overline{P_\mathrm{tx}}(\tau_\mathrm{tx}) = \rho_\mathrm{tx} \tau_\mathrm{tx}$. The slope $\rho_\mathrm{tx}$ depends linearly on the radio transmission parameters MCS and TXP.
 \item A component that accounts for the energy consumed in reception, which linearly grows with the airtime percentage $\tau_\mathrm{rx}$, i.e., $\overline{P_\mathrm{rx}}(\tau_\mathrm{rx}) = \rho_\mathrm{rx} \tau_\mathrm{rx}$. The slope $\rho_\mathrm{rx}$ depends linearly on the radio transmission parameter MCS.
 \item A new component, called \emph{generation cross-factor} or $\gamma_{\mathrm{xg}}$, that accounts for a per-frame energy processing toll in transmission, which linearly grows with the traffic rate $\lambda_\mathrm{g}$ generated, i.e., $\overline{P_\mathrm{xg}}(\lambda_\mathrm{g}) = \gamma_\mathrm{xg} \lambda_\mathrm{g}$. The slope $\gamma_\mathrm{xg}$ depends on the computing characteristics of the device.
 \item A new component, called \emph{reception cross-factor} or $\gamma_{\mathrm{xr}}$, that accounts for a per-frame energy processing toll in reception, which linearly grows with the traffic rate $\lambda_\mathrm{r}$ received, i.e., $\overline{P_\mathrm{xr}}(\lambda_\mathrm{r}) = \gamma_\mathrm{xr} \lambda_\mathrm{r}$. Likewise, the slope $\gamma_\mathrm{xr}$ depends on the computing characteristics of the device.
\end{itemize}

Therefore, the average power consumed $\overline{P}$ is a function of five device-dependent parameters ($\rho_i, \gamma_{\mathrm{x}i}$) and four traffic-dependent ones ($\tau_i, \lambda_i$).

\subsection{Energy efficiency analysis}

Putting together both models, we are now in a position to build a joint goodput-energy model for 802.11a DCF. Let us consider the average durations \eqref{Dsucc} and \eqref{Dfail}. Based on their expressions, we multiply the idle time ($\overline{\mathcal{D}}_\mathrm{wait}$, $\overline{T}_\mathrm{bkoff}$, $T_\mathrm{SIFS}$, $T_\mathrm{DIFS}$) by $\rho_\mathrm{id}$, the transmission time ($T_\mathrm{data}$) by $\rho_\mathrm{tx}$, and the reception time ($T_\mathrm{ACK}$) by $\rho_\mathrm{rx}$. The resulting expressions are the average energy consumed in a successful transmission $\mathcal{E}_{\mathrm{succ} \mid l, \hat{s}, \hat{m}}$ and the average energy wasted when a transmission fails $\mathcal{E}_{\mathrm{fail} \mid l, \hat{s}, \hat{m}}$:
\begin{align}
 \mathcal{E}_{\mathrm{succ} \mid l, \hat{s}, \hat{m}} = &\sum_{n=1}^{n_\mathrm{max}} \Pr[n \mathrm{~succ} \mid l, \hat{s}, \hat{m}] \cdot \biggl\lbrace \sum_{i=2}^{n_\mathrm{max}} \left[\rho_\mathrm{id}\overline{T}_\mathrm{bkoff}(i)\right.\biggr. \nonumber\\
 &+ \left.\rho_\mathrm{tx}T_\mathrm{data}(l, m_i) + \rho_\mathrm{id}\overline{\mathcal{D}}_\mathrm{wait}(i)\right] \nonumber\\
 &+ \rho_\mathrm{id}\overline{T}_\mathrm{bkoff}(1) + \rho_\mathrm{tx}T_\mathrm{data}(l, m_1) + \rho_\mathrm{id}T_\mathrm{SIFS} \nonumber\\
 &+ \biggl.\rho_\mathrm{rx}T_\mathrm{ACK}(m'_n) + \rho_\mathrm{id}T_\mathrm{DIFS} \biggr\rbrace
\end{align}
\begin{align}
 \mathcal{E}_{\mathrm{fail} \mid l, \hat{s}, \hat{m}} = &\sum_{i=1}^{n_\mathrm{max}} \left[\rho_\mathrm{id}\overline{T}_\mathrm{bkoff}(i)\right. \\
 &+ \left.\rho_\mathrm{tx}T_\mathrm{data}(l, m_i) + \rho_\mathrm{id}\overline{\mathcal{D}}_\mathrm{wait}(i+1)\right] \nonumber
\end{align}

Then, by analogy with \eqref{Ddata}, the \emph{expected energy consumed per frame transmitted}, $\e{\mathcal{E}_\mathrm{data}}$, can be written as follows:
\begin{align}\label{energyperframe}
 \e{\mathcal{E}_\mathrm{data}} = \gamma_\mathrm{xg} + \left(1 - \Pr[\mathrm{succ} \mid l, \hat{s}, \hat{m}]\right) \cdot \mathcal{E}_{\mathrm{fail} \mid l, \hat{s}, \hat{m}} \\
 + \Pr[\mathrm{succ} \mid l, \hat{s}, \hat{m}] \cdot \mathcal{E}_{\mathrm{succ} \mid l, \hat{s}, \hat{m}} \nonumber
\end{align}

It is noteworthy that the receiving cross-factor does not appear in this expression because ACKs (acknowledgements) are processed in the network card exclusively, and thus its processing toll is negligible.

Finally, we define the \emph{expected effective energy efficiency} $\mu$ as the ratio between the expected delivered data payload and the expected energy consumed per frame, which can be expressed in \emph{bits per Joule} (bpJ):
\begin{align}\label{efficiency}
 \mu(l, \hat{s}, \hat{m}) = \frac{\e{\mathrm{data}}}{\e{\mathcal{E}_\mathrm{data}}}
\end{align}

\section{Numerical results}\label{sec:results}

Building on the joint model presented in the previous section, here we explore the relationship between optimal goodput and energy efficiency in 802.11a. More specifically, our objective is to understand the behaviour of the energy efficiency of a single spatial stream as the MCS and TXP change following our model to meet the optimal goodput.

\subsection{Optimal goodput}

We note that the main goal of RA, generally, is to maximise the effective goodput that a station can achieve by varying the parameters of the interface. In terms of the model discussed in the previous section, a rate adaptation algorithm would aspire to fit the following curve:
\begin{align}\label{maxgoodput}
 \max{\mathcal{G}(l, \hat{s}, \hat{m})}
\end{align}

\begin{figure}[t]
	\centering
	\includegraphics[width=\linewidth]{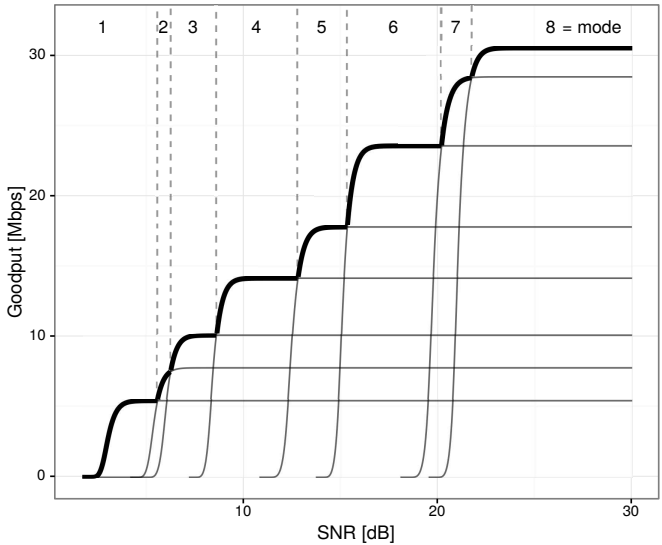}
	\caption{Optimal goodput (bold envelope) as a function of SNR.}
	\label{fig:maxgoodput}
\end{figure}

We provide the numerical results for this goodput maximisation problem in Fig.~\ref{fig:maxgoodput}, which are in good agreement with those obtained in \cite{Qiao2002}. For the sake of simplicity but without loss of generality we fix $l=1500$ octets and $n_\mathrm{max}=7$ retries, and assume that the channel conditions and the transmission strategy are constant across retries ($\hat{s}=\{s_1, \ldots, s_1\}$ and $\hat{m}=\{m_1, \ldots, m_1\}$).

Fig.~\ref{fig:maxgoodput} illustrates which mode (see Table~\ref{tab:modes}) is optimal in terms of goodput, given an SNR level. We next address the question of whether this optimisation is aligned with energy efficiency maximisation.

\begin{table}[t]
	\renewcommand{\arraystretch}{1.3}
	\footnotesize
	\caption{Modes of the IEEE 802.11a PHY}
	\label{tab:modes}
	\centering

	\begin{tabular}{r|rrrrrrrr}
	 \hline
	 Mode Index & 1 & 2 & 3 & 4 & 5 & 6 & 7 & 8 \\
	 \hline
	 MCS (Mbps) & 6 & 9 & 12 & 18 & 24 & 36 & 48 & 54 \\
	 \hline
	 \end{tabular}
\end{table}

\subsection{Extension of the energy parametrisation}

\begin{table*}[t]
	\renewcommand{\arraystretch}{1.3}
	\footnotesize
	\caption{Linear Regressions}
	\label{tab:regressions_tx}
	\centering
	\begin{tabular}{r|lll|c|ll|c}
		\hline
		\multirow{2}{*}{Device} & \multicolumn{4}{c|}{$\rho_\mathrm{tx}$ model estimates} & \multicolumn{3}{c}{$\rho_\mathrm{rx}$ model estimates} \\
		\cline{2-8}
		& $\alpha_0$ [W] & $\alpha_1$ [W/Mbps] & $\alpha_2$ [W/mW] & adj. $r^2$ & $\beta_0$ [W] & $\beta_1$ [W/Mbps] & adj. $r^2$ \\
		\hline
		HTC Legend & 0.354(14) & 0.0052(3) & 0.021(3) & 0.97 & 0.013(3) & 0.00643(11) & $>$0.99 \\
		Linksys WRT54G & 0.540(12) & 0.0028(2) & 0.075(3) & 0.98 & 0.14(2) & 0.0130(7) & 0.96 \\
		Raspberry Pi & 0.478(19) & 0.0008(4) & 0.044(5) & 0.88 & -0.0062(14) & 0.00146(5) & 0.98 \\
		Galaxy Note 10.1 & 0.572(4) & 0.0017(1) & 0.0105(9) & 0.98 & 0.0409(10) & 0.00173(4) & 0.99 \\
		Soekris net4826-48 & 0.17(3) & 0.0170(6) & 0.101(7) & 0.99 & 0.010(8) &0.0237(3) & $>$0.99 \\
		\hline
	\end{tabular}
\end{table*}

\begin{figure*}[t]
	\centering
	\csubfloat[$\rho_\mathrm{tx}$ fit as a function of MCS and TXP.]{
		\includegraphics[width=\linewidth]{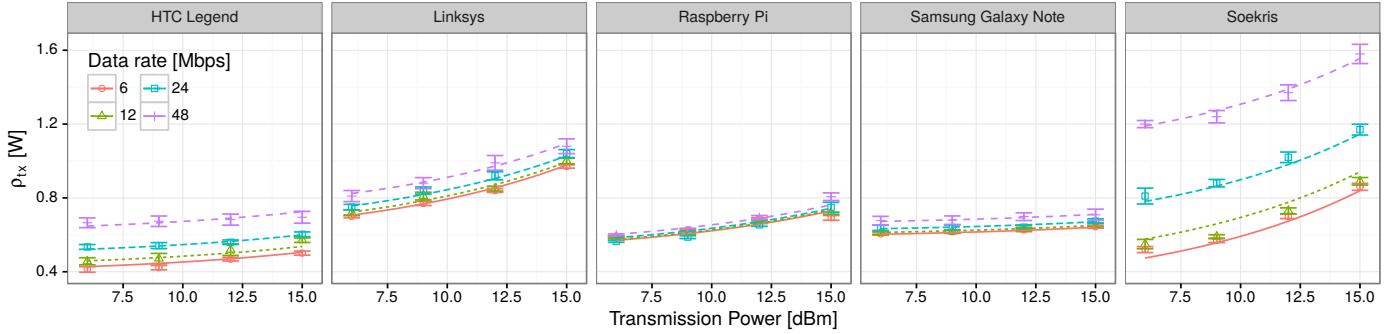}
		\label{fig:rho_tx}
	}\\
	\csubfloat[$\rho_\mathrm{rx}$ fit as a function of MCS.]{
		\includegraphics[width=\linewidth]{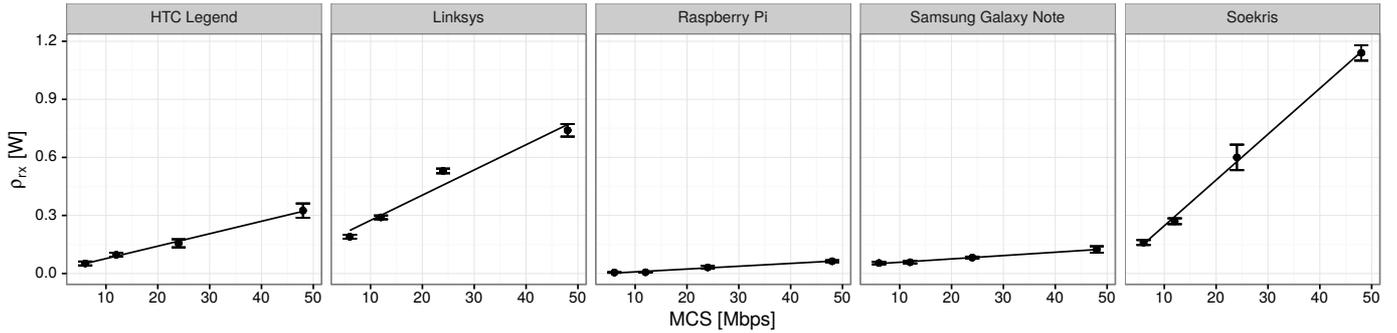}
		\label{fig:rho_rx}
	}
	\caption{Linear regressions.}
\end{figure*}

The next step is to delve into the energy consumption of wireless devices. \cite{Serrano2014} provides real measurements for five devices: three AP-like platforms (Linksys WRT54G, Raspberry Pi and Soekris net4826-48) and two hand-held devices (HTC Legend and Samsung Galaxy Note 10.1). Two of the four parameters needed are constant ($\rho_\mathrm{id}, \gamma_\mathrm{xg}$), and the other two ($\rho_\mathrm{tx}, \rho_\mathrm{rx}$) depend on the MCS and the TXP used. However, the characterisation done in \cite{Serrano2014} is performed for a subset of the MCS and TXP available, so we next detail how we extend the model to account for a larger set of operation parameters.

A detailed analysis of the numerical figures presented in \cite{Serrano2014} suggests that $\rho_\mathrm{rx}$ depends linearly on the MCS, and that $\rho_\mathrm{tx}$ depends linearly on the MCS and the TXP (in mW). Based on these observations, we define the following linear models:
\begin{align}
 \rho_\mathrm{tx} &= \alpha_0 + \alpha_1\cdot\mathrm{MCS} + \alpha_2\cdot\mathrm{TXP} \\
 \rho_\mathrm{rx} &= \beta_0 + \beta_1\cdot\mathrm{MCS}
\end{align}

The models are fed with the data reported in \cite{Serrano2014}, and the resulting fitting is illustrated in Figs.~\ref{fig:rho_tx} and \ref{fig:rho_rx}, while Table~\ref{tab:regressions_tx} collects the model estimates for each device (with errors between parentheses), as well as the adjusted r-squared. Since these linear models show a very good fit, they support the generation of synthetic data for the different MCS and TXP required.

\subsection{Energy consumption}

\begin{figure*}[t]
	\centering
	\includegraphics[width=\linewidth]{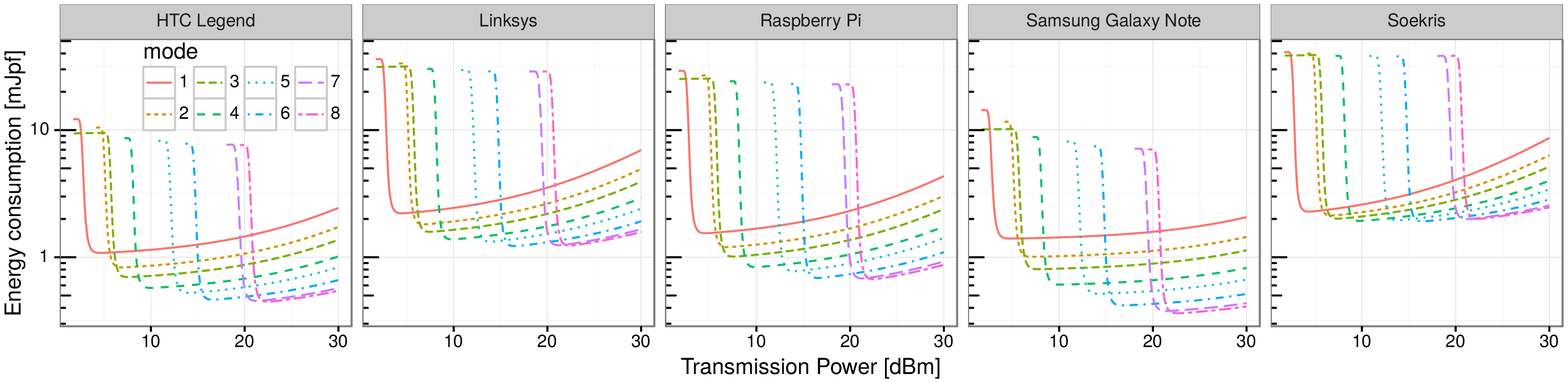}
	\caption{Expected energy consumption per frame in \emph{millijoules per frame} (mJpf) under fixed channel conditions.}
	\label{fig:consumption}
\end{figure*}

To compute the energy consumption using the above parametrisation, first we have to define the assumptions for the considered scenario. We assume for simplicity a device-to-device communication, with fixed and reciprocal channel conditions during a sufficient period of time (i.e., low or no mobility). As we have discussed before, our primary goal is to isolate MCS and TXP as variables of interest, but we must not forget that these are also reasonable assumptions in scenarios targeted by recent 802.11 standard developments (11ac, 11ad).

For instance, given channel state information from a receiver, the transmitter may decide to increase the TXP in order to increase the receiver's SNR (and thus the expected goodput), or to decrease it if the channel quality is high enough. Although the actual relationship between TXP and SNR depends on the specific channel model (e.g., distance, obstacles, noise), without loss of generality, we choose a noise floor of $N=-85$~dBm in an office scenario with a link distance of $d=18$~m in order to explore numerically the whole range of SNR while using reasonable values of TXP. The ITU model for indoor attenuation \cite{iturp1238-2015} gives a path loss of $L\approx 85$~dBm. Then, we can use \eqref{energyperframe} to obtain the expected energy consumed per frame and MCS mode, with TXP being directly related to the SNR level.

The results are reported in Fig.~\ref{fig:consumption}. As the figure illustrates, consumption first falls abruptly as the TXP increases for all modes, which is caused when the SNR reaches a sharp threshold level such that the number of retransmissions changes from 6 to 0 (i.e., no frame is discarded). From this threshold on, the consumption increases with TXP because, although the number of retransmissions is 0, the wireless interface consumes more power. We note that the actual value of the TXP when the consumption drops depends on the specifics of the scenario considered, but the qualitative conclusions hold for a variety of scenarios.

\subsection{Energy efficiency vs. optimal goodput}

\begin{figure}[t]
	\centering
	\includegraphics[width=\linewidth]{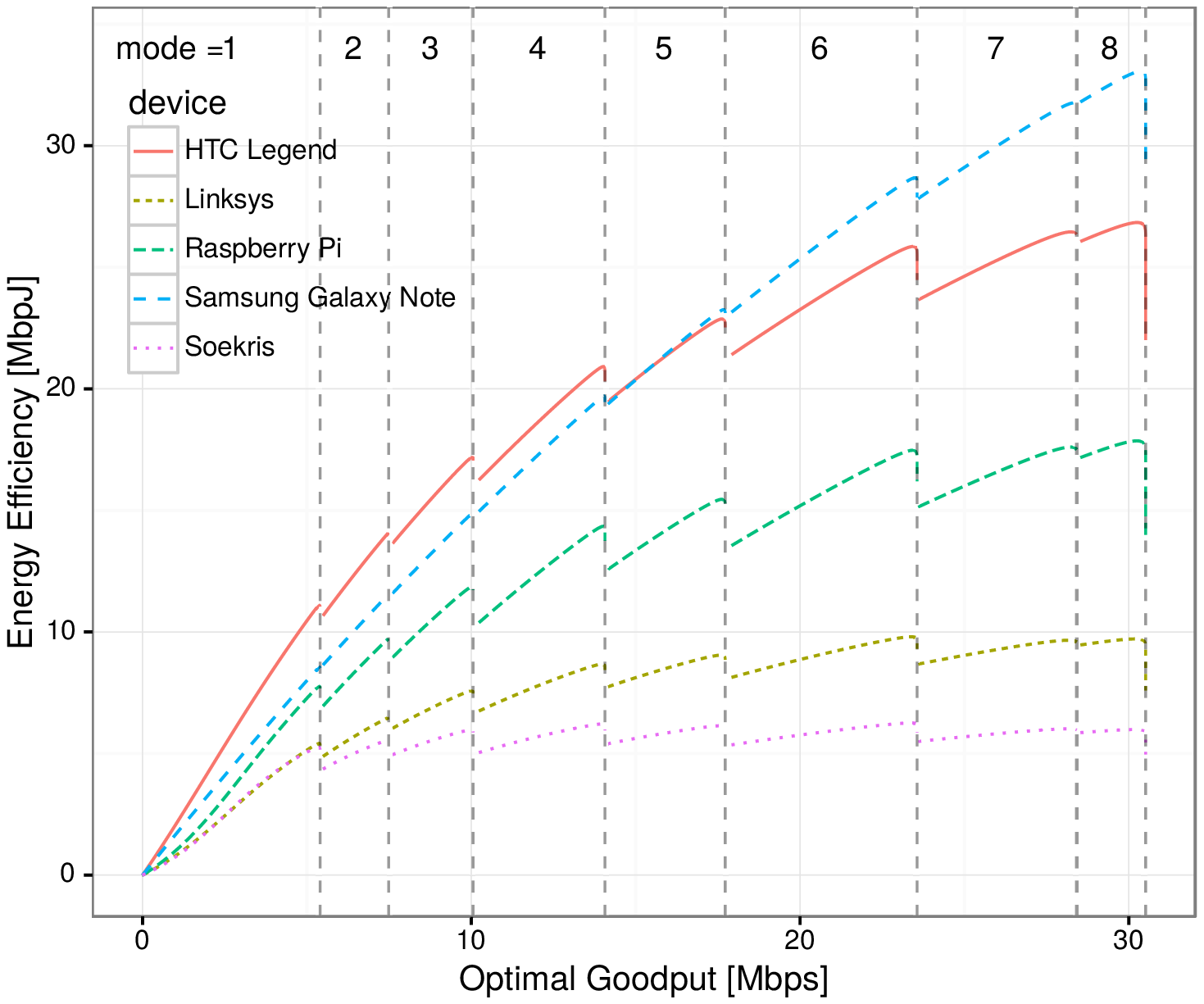}
	\caption{Energy efficiency vs. optimal goodput under fixed channel conditions.}
	\label{fig:efficiency-goodput}
\end{figure}

We can finally merge previous numerical analyses and confront energy efficiency, given by \eqref{efficiency}, and optimal goodput, given by \eqref{maxgoodput}, for all devices and under the aforementioned assumptions. To this aim, we plot in the same figure the energy efficiency for the configuration that maximises goodput given an SNR value vs. the obtained goodput, with the results being depicted in Fig.~\ref{fig:efficiency-goodput}. We next discuss the main findings from the figure.

First of all, we can see that the energy efficiency grows sub-linearly with the optimal goodput (the optimal goodput for each SNR value) in all cases. We may distinguish three different cases in terms of energy efficiency: high (Samsung Galaxy Note and HTC Legend), medium (Raspberry Pi) and low energy efficiency (Linksys and Soekris). Furthermore, for the case of the Soekris, we note that the ``central modes'' (namely, 4 and 5) are more efficient in their optimal region than the subsequent ones.

Another finding (more relevant perhaps) is that it becomes evident that increasing the goodput does not always improve the energy efficiency: there are more or less drastic leaps, depending on the device, between mode transitions. From the transmitter point of view, in the described scenario, this can be read as follows: we may increase the TXP to increase the SNR, but if the optimal goodput entails a mode transition, the energy efficiency may be affected.

As a conclusion, we have demonstrated that optimal goodput and energy efficiency do not go hand in hand, even in a single spatial stream, in 802.11. There is a trade-off in some circumstances that current rate adaptation algorithms cannot take into account, as they are oblivious to the energy consumption characteristic of the device.

\subsection{Sensitivity to energy parameter scaling}\label{sec:param-scaling}

We next explore how the different energy parameters affect the energy efficiency vs. optimal goodput relationship. For this purpose, we selected the Raspberry Pi curve from Fig.~\ref{fig:efficiency-goodput} (results are analogous with the other devices) and we scale up and down, one at a time, the four energy parameters $\rho_\mathrm{id}$, $\rho_\mathrm{tx}$, $\rho_\mathrm{rx}$, and $\gamma_\mathrm{xg}$. The scaling up and down is done by multiplying and dividing by 3, respectively, the numerical value of the considered parameter. One of the first results is that the impact of $\rho_\mathrm{rx}$ is negligible, a result somehow expected as the cost of receiving the ACK is practically zero. From this point on, we do not consider further this parameter.

\begin{figure}[t]
	\centering
	\subfloat[Overall effect.]{
		\includegraphics[width=\linewidth]{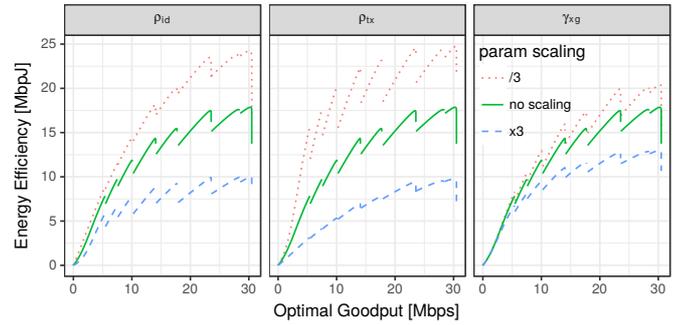}
		\label{fig:param-scaling1}
	}\\
	\subfloat[Impact on mode transitions.]{
		\includegraphics[width=\linewidth]{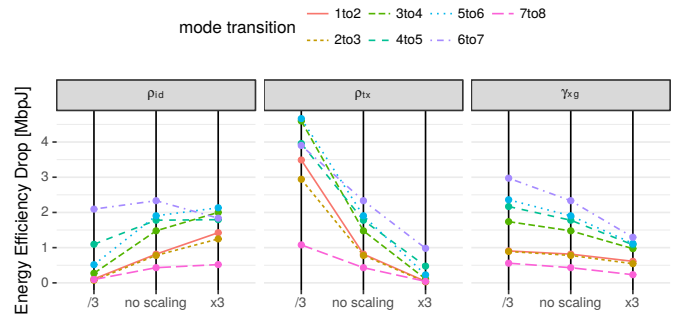}
		\label{fig:param-scaling2}
	}
	\caption{Impact of energy parameter scaling on the energy efficiency.}
\end{figure}

We show in Fig.~\ref{fig:param-scaling1} the overall effect of this parameter scaling. The solid line represents the base case with no scaling (same curve as in Fig.~\ref{fig:efficiency-goodput}), and in dashed and dotted lines the corresponding parameter was multiplied or divided by a factor of 3, respectively. As expected, larger parameters contribute to lower the overall energy efficiency. However, the impact on the energy efficiency drops between mode transitions is far from being obvious, as in some cases transitions are more subtle while in others they become more abrupt.

To delve into these transitions, we illustrate in Fig.~\ref{fig:param-scaling2} the ``drop'' in energy efficiency when changing between modes. As it can be seen, the cross-factor $\gamma_\mathrm{xg}$ is the less sensitive parameter of the three, because its overall effect is limited and, more importantly, it scales all the leaps between mode transitions homogeneously. This means that a higher or lower cross-factor, which resides almost entirely in the device and not in the wireless card, does not alter the energy efficiency vs. optimal goodput relationship (note that this parameter results in a constant term in \eqref{energyperframe}). Thus, the cross-factor is not relevant from the RA-TPC point of view, and energy-aware RA-TPC algorithms can be implemented by leveraging energy parameters local to the wireless card.

On the other hand, $\rho_\mathrm{id}$ and $\rho_\mathrm{tx}$ have a larger overall effect, plus an inhomogeneous and, in general, opposite impact on mode transitions. While a larger $\rho_\mathrm{id}$ contributes to larger leaps, for the case of $\rho_\mathrm{tx}$, the larger energy efficiency drops occur with smaller values of that parameter. Still, the reason behind this behaviour is the same for both cases: the wireless card spends more time in idle (and less time transmitting) when a transition to the next mode occurs, which has a higher data rate.

This effect is also evident if we compare the Samsung Galaxy Note and the HTC Legend curves in Fig.~\ref{fig:efficiency-goodput}. Both devices have $\rho_\mathrm{id}$ and $\rho_\mathrm{tx}$ in the same order of magnitude, but the HTC Legend has a larger $\rho_\mathrm{id}$ and a smaller $\rho_\mathrm{tx}$. The combined outcome is a more dramatic sub-linear behaviour and an increased energy efficiency drop between mode transitions.

\subsection{Discussion}\label{sec:conservativeness}

We have seen that the energy efficiency vs. optimal goodput relationship shows a signature ``sawtooth'' pattern when RA and TPC are considered for a single 802.11 spatial stream. This sawtooth shape presents a growing trend in the central part of each mode, but there are energy efficiency drops between mode transitions, which conceal a trade-off.

Parameter scaling has diverse effects on the final consumption signature, but overall, the qualitative behaviour (i.e., the shape) remains the same. The cross-factor produces an homogeneous scaling of the sawtooth. Thus, a first conclusion is that the trade-off depends on the energy parameters local to the wireless card, which means that a properly designed energy-aware RA-TPC algorithm can be device-agnostic.

Moreover, an energy-aware RA-TCP algorithm may also be card-agnostic. This is because the inefficiencies are always constrained at mode transitions, which are exactly the points at which RA-TPC algorithms take decisions. Therefore, there is no need of knowing the exact energy parametrisation, nor the instantaneous power consumption of the wireless card, in order to take energy-efficient decisions.

An RA-TPC algorithm moves along the sawtooth shapes of Fig.~\ref{fig:efficiency-goodput} in two directions, namely, ``up'' (towards higher throughput) and ``down'' (towards lower throughput). In this way, an algorithm requires different policies to make a decision: ($i$)~the \emph{upwards} policy, in which mode transitions take place by increasing MCS and TXP (to achieve more goodput), and ($ii$)~the \emph{downwards} policy, in which mode transitions take place by decreasing MCS and TXP.

\begin{enumerate}[\itshape(i)]
 \item In the \emph{upwards} direction, a sensitive policy would be to remain in the left side of the leaps, to prevent falling into the efficiency gaps, until the link is good enough to move to a higher MCS with at least the same efficiency. An heuristic for the \emph{upwards} policy may be the following: whenever an algorithm chooses a higher MCS, it may hold the decision for some time and, if it persists, then trigger the MCS change (however, if this delay is too long, the algorithm will incur in inefficiencies, too).
 \item In the \emph{downwards} direction, a sensitive policy would be to try to reach the left side of the leaps as soon as possible. However, it should be noted that this \emph{downwards} policy is much more challenging, as it implies predicting quality drops to trigger early MCS/TXP changes.
\end{enumerate}

In summary, our results suggest that one of the key points of an energy-aware RA-TPC algorithm is the management of mode transitions. A good algorithm should be \emph{conservative} at mode transitions, in the sense that it should prefer a lower MCS and TXP until a higher MCS can be guaranteed. 

\section{Experimental validation}\label{sec:experiments}

This section is devoted to experimentally validate the results from the numerical analysis and, therefore, the resulting conclusions. To this aim, we describe our experimental setup and the validation procedure, first specifying the methodology and then the results achieved.

\subsection{Experimental setup}

\begin{figure}[t]
	\centering
	\includegraphics[width=\linewidth]{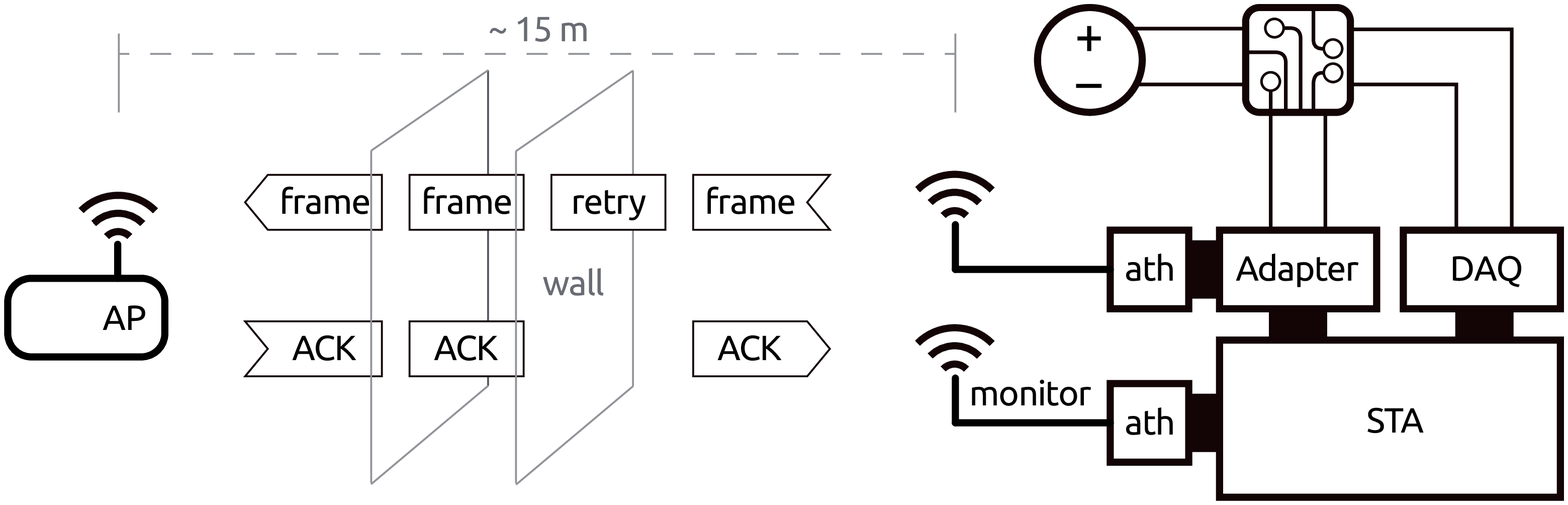}
	\caption{Experimental setup.}
	\label{fig:testbed}
\end{figure}

We deployed the testbed illustrated in Fig.~\ref{fig:testbed}, which consists of a station (STA) transmitting evenly-spaced maximum-sized UDP packets to an access point (AP). The AP is an x86-based Alix6f2 board with a Mini PCI Qualcomm Atheros AR9220 wireless network adapter, running Voyage Linux with kernel version 3.16.7 and the \texttt{ath9k} driver. The STA is a desktop PC with a Mini PCI Express Qualcomm Atheros QCA9880 wireless network adapter, running Fedora Linux 23 with kernel version 4.2.5 and the \texttt{ath10k} driver. We also installed at the STA a Mini PCI Qualcomm Atheros AR9220 wireless network adapter to monitor the wireless channel.

The QCA9880 card is connected to the PC through a \emph{x1 PCI Express to Mini PCI Express} adapter from Amfeltec. This adapter connects the PCI bus' data channels to the host and provides an ATX port so that the wireless card can be supplied by an external power source. The power supply is a Keithley 2304A DC Power Supply, and it powers the wireless card through an \emph{ad-hoc} measurement circuit that extracts the voltage and converts the current with a high-precision sensing resistor and amplifier. These signals are measured using a National Instruments PCI-6289 multifunction data acquisition (DAQ) device, which is also connected to the STA. Thanks to this configuration, the STA can simultaneously measure the instant power consumed by the QCA9880 card,\footnote{Following the discussion on Section~\ref{sec:param-scaling} the device's cross-factor is not involved in the trade-off, thus we will expect to reproduce it by measuring the wireless interface alone.} and the goodput achieved.

As Fig.~\ref{fig:testbed} illustrates, the STA is located in an office space and the AP is placed in a laboratory 15~m away, and transmitted frames have to transverse two thin brick walls. The wireless card uses only one antenna and a practically-empty channel in the 5-GHz band. Throughout the experiments, the STA is constantly backlogged with data to send to the AP, and measures the throughput obtained by counting the number of received acknowledgements (ACKs).

\subsection{Methodology and results}

\begin{figure}[t]
	\centering
	\includegraphics[width=\linewidth]{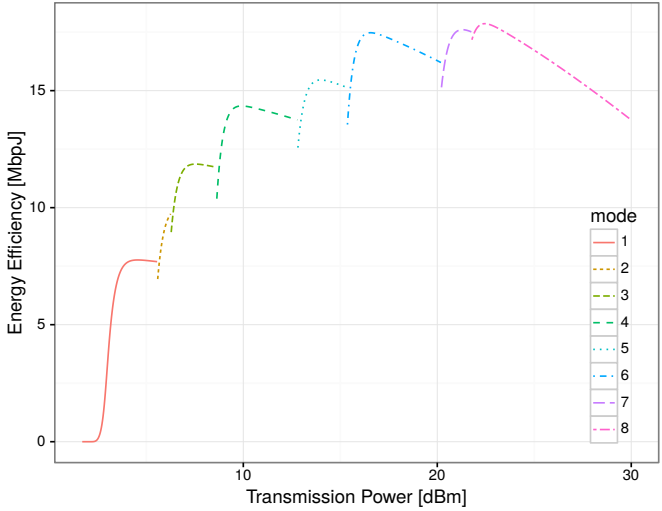}
	\caption{Energy Efficiency vs. Transmission Power under fixed channel conditions for the Raspberry Pi case.}
	\label{fig:efficiency-txp}
\end{figure}

\begin{figure}[t]
	\centering
	\includegraphics[width=\linewidth]{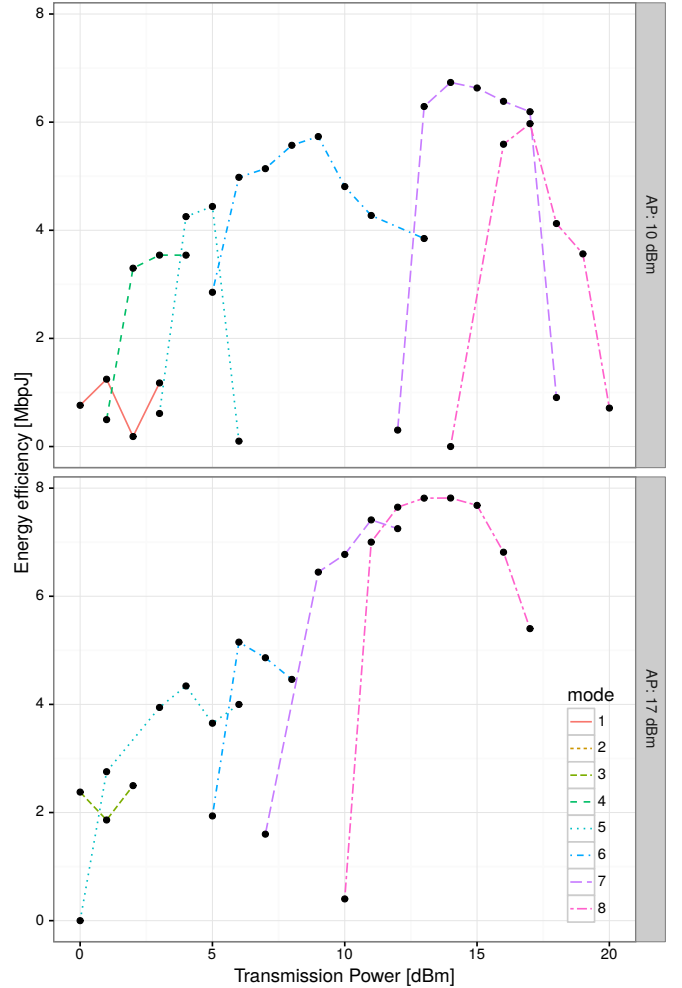}
	\caption{Experimental study of Fig.~\ref{fig:efficiency-txp} for two AP configurations.}
	\label{fig:efficiency-txp-exp}
\end{figure}

In order to validate our results, our aim is to replicate the qualitative behaviour of Fig.~\ref{fig:efficiency-goodput}, in which there are energy efficiency ``drops'' as the optimal goodput increases. However, in our experimental setting, channel conditions are not controllable, which introduces a notable variability in the results as it affects both the $x$-axis (goodput) and the $y$-axis (energy efficiency). To reduce the impact of this variability, we decided to change the variable in the $x$-axis from the optimal goodput to the transmission power ---a variable that is directly configured in the wireless card---. In this way, the qualitative behaviour to replicate is the one illustrated in Fig.~\ref{fig:efficiency-txp}, where we can still identify the performance ``drops'' causing the loss in energy efficiency.

Building on Fig.~\ref{fig:efficiency-txp}, we perform a sweep through all available combinations of MCS (see Table~\ref{tab:modes}) and TXP.\footnote{The model explores a range between 0 and 30 dBm to get the big picture, but this particular wireless card only allows us to sweep from 0 to 20 dBm.} Furthermore, we also tested two different configurations of the AP's TXP at different times of the day, to confirm that this qualitative behaviour is still present under different channel conditions. For each configuration, we performed 2-second experiments in which we measure the total bytes successfully sent and the energy consumed by the QCA9880 card with sub-microsecond precision, and we compute the energy efficiency achieved for each experiment.

The results are shown in Fig.~\ref{fig:efficiency-txp-exp}. Each graph corresponds to a different TXP value configured at the AP, and depicts a single run (note that we performed several runs throughout the day and found no major qualitative differences across them). Each line type represents the STA's mode that achieved the highest goodput for each TXP interval, therefore in some cases low modes do not appear because a higher mode achieved a higher goodput. Despite the inherent experimental difficulties, namely, the low granularity of 1-dBm steps and the random variability of the channel, the experimental results validate the analytical ones, as the qualitative behaviour of both graphs follows the one illustrated in Fig.~\ref{fig:efficiency-txp}. In particular, the performance ``drops'' of each dominant mode can be clearly observed (especially for the 36, 48 and 54 Mbps MCSs) despite the variability in the results.

\section{On the performance of RA-TPC algorithms}\label{sec:simulations}

So far, we have demonstrated through numerical analysis, and validated experimentally, the existence of a trade-off between two competing performance figures, namely, goodput and energy efficiency. This issue arises even for a single spatial stream in absence of interference.  Furthermore, we have discussed in Section~\ref{sec:conservativeness} some ideas about the kind of mechanisms that energy-aware RA-TPC algorithms may incorporate, to leverage the behaviour that we have identified in our analysis in these so-called mode transitions. \emph{In nuce}, the algorithms should be \emph{conservative} during these transitions.

During that discussion, we neglected the challenge of estimating channel conditions. In practice, any RA-TPC algorithm has imperfect channel knowledge, and therefore will adapt to changing conditions in a suboptimal way. In this section, we will analyse and compare the performance of several representative existing RA algorithms, which also incorporate TPC, to confirm if the \emph{conservativeness} in such decisions may have a positive impact on the achieved performance.

\subsection{Considered RA-TPC algorithms}

If we take a look at the actual operation of WiFi networks, the Minstrel algorithm \cite{minstrel}, which was integrated into the Linux kernel, has become the \emph{de facto} standard due to its relatively good performance and robustness. However, Minstrel does not consider TPC and, in consequence, there is no TPC in today's WiFi deployments. Moreover, despite some promising proposals have been presented in the literature, there are very few of them implemented, although there are some ongoing efforts such as the work by the authors of \emph{Minstrel-Piano} \cite{huehn2012}, who are pushing to release an enhanced version of the latter for the Linux kernel with the goal of promoting it upstream.\footnote{\url{https://github.com/thuehn/Minstrel-Blues}}

As stated before, RA is a very prolific research line in the literature, but the main corpus is dedicated to the MCS adjustment without taking into account the TXP \cite{biaz2008}. There is some work considering TPC, but the motivation is typically the performance degradation due to network densification, and the aim is interference mitigation \cite{richart2015} and not energy efficiency. Given that we are interested in assessing RA implementations with TPC support, we consider only open-source algorithms that can be tested using the NS3 Network Simulator. After a thorough analysis of the literature, we consider the following set of algorithms: 

\begin{itemize}
 \item \emph{Power-controlled Auto Rate Fallback} (PARF) \cite{Akella:2005}, which is based on \emph{Auto Rate Fallback} (ARF) \cite{kamerman1997wavelan}, one of the earliest RA schemes for 802.11. ARF rate adaptation is based on the frame loss ratio. It tunes the MCS in a very straightforward and intuitive way. The procedure starts with the lowest possible MCS. Then, if either a timer expires or the number of consecutive successful transmissions reach a threshold, the MCS is increased and the timer is reset. The MCS is decreased if either the first transmission at a new rate fails or two consecutive transmissions fail. PARF builds on ARF and tries to reduce the TXP if there is no loss until a minimum threshold is reached or until transmissions start to fail. If transmission fails persist, the TXP is increased.
 \item \emph{Minstrel-Piano} (MP) \cite{huehn2012} is based on Minstrel \cite{minstrel}. Minstrel performs per-frame rate adaptation based on throughput. It randomly probes the MCS space and computes an exponential weighted moving average (EWMA) on the transmission probability for each rate, in order to keep a long-term history of the channel state. As the previous algorithm, MP adds TPC without interfering with the normal operation of Minstrel. It incorporates to the TPC the same concepts and techniques than Minstrel uses for the MCS adjustment, i.e., it tries to learn the impact of the TXP on the achieved throughput.
 \item \emph{Robust Rate and Power Adaptation Algorithm} (RRPAA) and \emph{Power, Rate and Carrier-Sense Control} (PRCS) \cite{richart2015}, which are based on \emph{Robust Rate Adaptation Algorithm} (RRAA) \cite{Wong2006}. RRAA consists of two functional blocks, namely, rate adaptation and collisions elimination. It performs rate adaptation based on loss ratio estimation over short windows, and reduces collisions with a RTS-based strategy. The procedure starts at the maximum MCS. The loss ratio for each window of transmissions is available for rate adjustment in the next window. There are two thresholds involved in this adjustment: if the loss ratio is below both of them, the MCS is increased; if it is above, the MCS is decreased; and if it is in between, the MCS remains unchanged. RRPAA and PRCS build on this and try to use the lowest possible TXP without degrading the throughput. For this purpose, they firstly find the best MCS at the maximum TXP and, from there, they jointly adjust the MCS and TXP for each window based on a similar thresholding system. RRPAA and PRCS are very similar and only differ in implementation details.
\end{itemize}

Based on their behaviour, these algorithms can be classified into three distinct classes. First of all, MP is the most aggressive technique, given that it constantly samples the whole MCS/TXP space searching for the best possible configuration. On the opposite end, RRPAA and PRCS do not sample the whole MCS/TXP space. Instead, they are based on a windowed estimation of the loss ratio, which makes the MCS/TXP transitions much lazier. Finally, PARF falls in between, as it changes the MCS/TXP to the next available proactively if a number of transmissions are successful, but it falls back to the previous one if the new one fails. In practice, this may result in some instability during transitions.

\subsection{Scenario}

\begin{figure}[t]
	\centering
	\includegraphics[width=\linewidth]{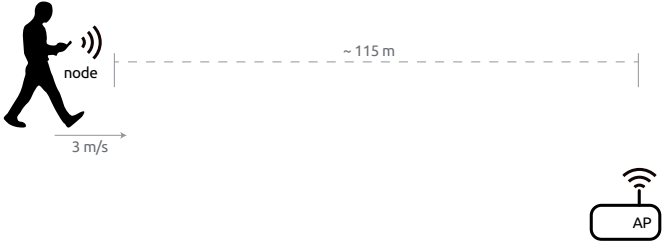}
	\caption{Simulation scenario.}
	\label{fig:simulation}
\end{figure}

This evaluation is publicly available,\footnote{\url{https://github.com/Enchufa2/ns-3-dev-git}} and builds upon the code provided by Richart \emph{et al.} in \cite{richart2015}.\footnote{\url{https://github.com/mrichart/ns-3-dev-git}}. We assessed the proposed algorithms in the toy scenario depicted in Fig.~\ref{fig:simulation}. It consists of a single access point (AP) and a single mobile node connected to this AP configured with the 802.11a PHY. The mobile node at the farthest distance at which is able to communicate at the lowest possible rate (6 Mbps) and highest TXP (17 dBm), and then it moves at constant speed towards the AP. The simulation stops when the node is directly in front of the AP and it is able to communicate at the highest possible rate (54 Mbps) and lowest TXP (0 dBm). This way, we sweep through all mode transitions available.

For the whole simulation, the AP tries to constantly saturate the channel by sending full-size UDP packets to the node. Every transmission attempt is monitored, as well as every successful transmission. The first part allows us to compute the transmission time, while the latter allows us to compute the reception time (of the ACKs) and the goodput achieved.

The simulation model assembles the power model (\ref{eq:power}) with the parametrisation previously made (see Table~\ref{tab:regressions_tx}) for all the devices considered in Section~\ref{sec:results}: HTC Legend, Linksys WRT54G, Raspberry Pi, Samsung Galaxy Note 10.1 and Soekris net4826-48. Thus, the total energy consumed is computed for all the devices and each run using the computed transmission time, reception time and idle time. The beacons are ignored and considered as idle time.

We set up one simulation for each algorithm (PARF, MP, PRCS, RRPAA) with a fixed seed, and perform 10 independent runs for each simulation. We use boxplots for the results unless otherwise mentioned.

\subsection{Results}

\begin{figure}[t]
	\centering
	\includegraphics[width=\linewidth]{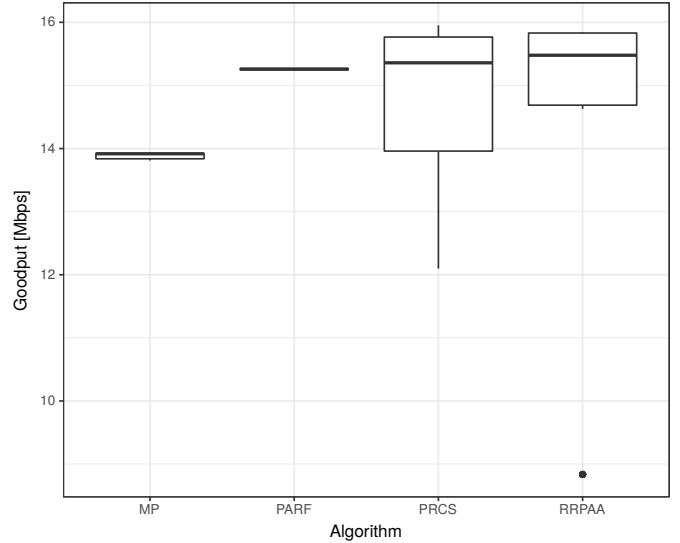}
	\caption{Goodput achieved per simulated algorithm.}
	\label{fig:ns3-goodput}
\end{figure}

\begin{figure*}[!h]
	\centering
	\includegraphics[width=\linewidth]{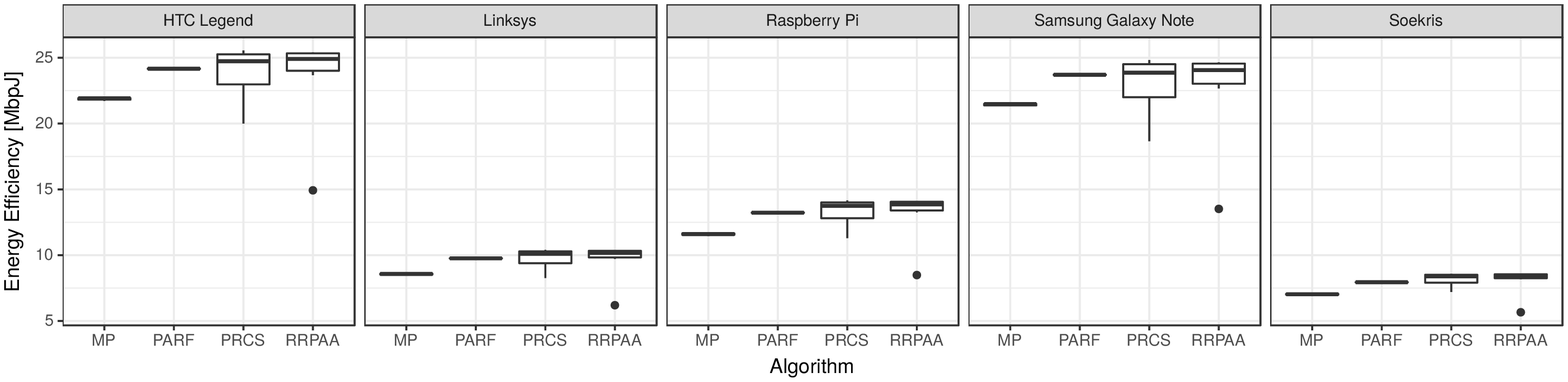}
	\caption{Energy efficiency achieved per simulated algorithm and device.}
	\label{fig:ns3-efficiency}

	\centering
	\csubfloat[MCS vs. simulation time.]{
		\includegraphics[width=\linewidth]{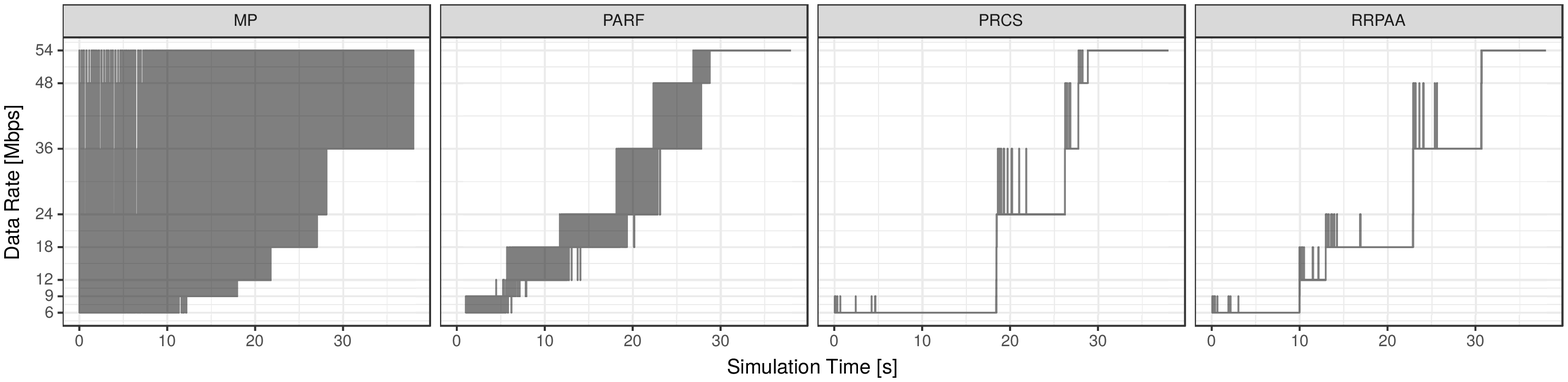}
		\label{fig:ns3-rate}
	}\\
	\csubfloat[TXP vs. simulation time.]{
		\includegraphics[width=\linewidth]{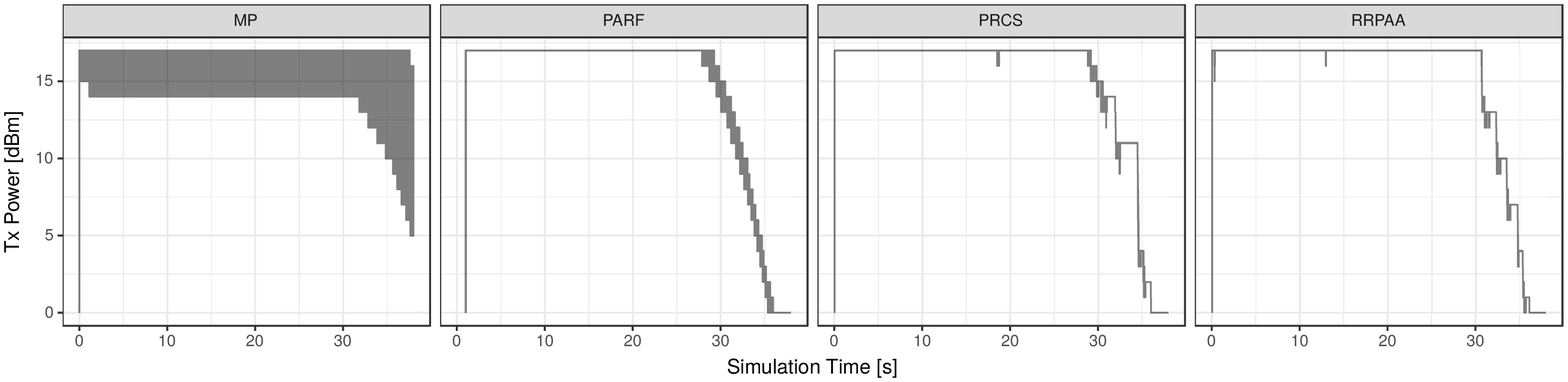}
		\label{fig:ns3-txpower}
	}
	\caption{MCS and TXP evolution per algorithm for a selected run.}
\end{figure*}

We first analyse the goodput achieved per each algorithm, which are depicted in Fig.~\ref{fig:ns3-goodput}. The median of the average goodput across several runs for RRPAA is the highest, followed by PRCS, PARF and MP. PRCS and RRPAA, which are very similar mechanisms, show a higher variability across replications compared to PARF and MP, which have little dispersion.

Fig.~\ref{fig:ns3-efficiency} shows the energy efficiency achieved per algorithm, computed for all the devices presented in Section~\ref{sec:results}. As expected, the numerical values of the energy efficiency achieved are different across devices, but the relative performance is essentially the same, as in the previous case. Indeed, the efficiency follows the pattern seen in Fig.~\ref{fig:ns3-goodput}: RRPAA results the most energy efficient in our scenario, followed by PRCS, PARF and MP. PRCS and RRPAA exhibit the same variability across replications as in the case of goodput, which is particularly notable for the most efficient devices, i.e., the HTC Legend and the Samsung Galaxy Note.

\subsection{Discussion}

In order to shed some light into the reasons behind the differences in performance, Figs.~\ref{fig:ns3-rate} and \ref{fig:ns3-txpower} show the behaviour of each algorithm throughout the simulation time for one run, showing the evolution of the MCS and TXP chosen by each algorithm, respectively. Here, we can clearly differentiate that there are two kinds of behaviour: while MP and PARF are constantly sampling other MCSs and TXPs, PRCS and RRPAA are much more conservative in that sense, and tend to keep the same configuration for longer periods of time. 
  
MP randomly explores the whole MCS/TXP space above a basic \emph{guaranteed} value, and this is the explanation for the apparently uniformly greyed zone. Also, this aggressive approach is clearly a disadvantage in the considered toy scenario (deterministic walk, one-to-one, no obstacles), and this is why the achieved goodput in Fig.~\ref{fig:ns3-goodput} is slightly smaller than the one achieved by the others. PARF, on its part, only explores the immediately higher MCS/TXP, which leads to a higher goodput and efficiency.

On the other hand, PRCS and RRPAA sampling is much more sparse in time. As a consequence, Figs.~\ref{fig:ns3-rate} and \ref{fig:ns3-txpower} are much more different across replications, leading to the high variability shown in Fig.~\ref{fig:ns3-goodput} compared to MP and PARF.
  
In terms of TXP, all the algorithms exhibit a similar \emph{aggressiveness}, in the sense that they use a high TXP value in general. Indeed, as Fig.~\ref{fig:ns3-txpower} shows, the TXP is the highest possible until the very end of the simulation, when the STA is very close to the AP. This is the cause for the high correlation between Figs.~\ref{fig:ns3-goodput} and \ref{fig:ns3-efficiency}.
  
A noteworthy characteristic of PRCS and RRPAA is that, in general, they \emph{delay} the MCS change decision, as depicted in Fig.~\ref{fig:ns3-rate}. Most of the times, they do not even use the whole space of MCS available, unlike MP and PARF. Because of this, they tend to achieve the best goodput and energy efficiency.

\subsection{Conservativeness at mode transitions}

\begin{figure*}[t]
	\centering
	\includegraphics[width=\linewidth]{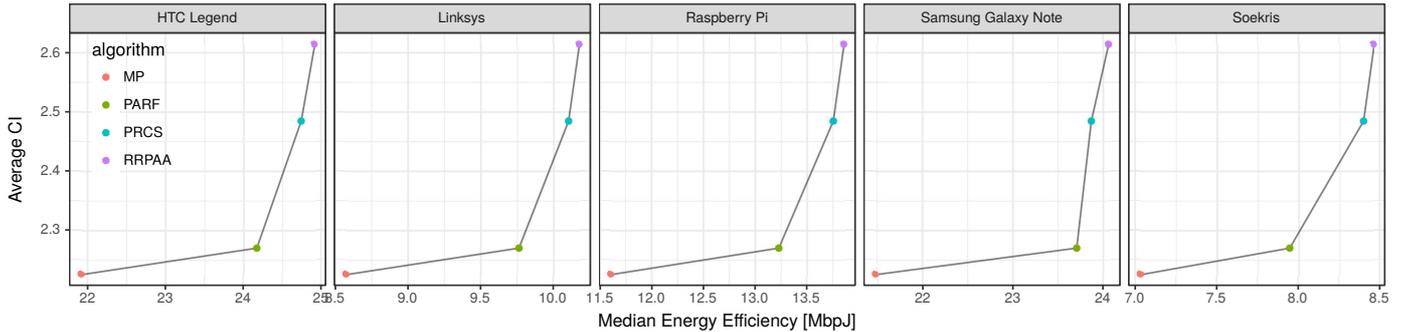}
	\caption{Relationship between Conservativeness Index (tendency to select lower MCS and TXP) and energy effienciency per simulated device.}
	\label{fig:ns3-conservativeness}
\end{figure*}

Building on the concept of \emph{conservativeness} developed in Section~\ref{sec:conservativeness} (i.e., the tendency to select a lower MCS/TXP in the transition regions), we explore whether there is any correlation of with the energy efficiency achieved by a certain algorithm and this tendency. For that purpose, we first define a proper metric.

In the first place, we define the \emph{normalised average MCS} as the area under the curve in Fig.~\ref{fig:ns3-rate} normalised by the total simulation time and the maximum MCS:
\begin{align}
 \widehat{\mathrm{MCS}} = \frac{1}{\max(\mathrm{MCS}) \cdot t_\mathrm{sim}}\int_0^{t_\mathrm{sim}} \mathrm{MCS}(t)dt
\end{align}

\noindent where $t_\mathrm{sim}$ is the simulation time and $\max(\mathrm{MCS})$ is 54 Mbps in our case. The same concept can be applied to the TXP:
\begin{align}
 \widehat{\mathrm{TXP}} = \frac{1}{\max(\mathrm{TXP}) \cdot t_\mathrm{sim}}\int_0^{t_\mathrm{sim}} \mathrm{TXP}(t)dt
\end{align}

\noindent where $\max(\mathrm{TXP})$ is 17 dBm in our case. Both $\widehat{\mathrm{MCS}}$ and $\widehat{\mathrm{TXP}}$ are unitless scores between 0 and 1, and lower values mean a more conservative algorithm. Therefore, we can define a \emph{Conservativeness Index} (CI) as the inverse of the product of both scores:
\begin{align}
 \mathrm{CI} = \frac{1}{\widehat{\mathrm{MCS}} \cdot \widehat{\mathrm{TXP}}}
\end{align}

\noindent where $\mathrm{CI}>1$.\footnote{It must be taken into account that the CI is not suitable for comparing \emph{any} algorithm. For instance, in an extreme case, an ``algorithm'' could select 6 Mbps and 0 dBm always, resulting in a very low CI, but a very bad performance at the same time. The CI should only be used for comparing similarly performant algorithms, as it is the case in our study given the results shown in Figs.~\ref{fig:ns3-goodput} and \ref{fig:ns3-efficiency}.}

We computed the CI for each device and run, and the final results are depicted in Fig.~\ref{fig:ns3-conservativeness} as the average CI across different runs vs. the median energy efficiency in Fig.~\ref{fig:ns3-efficiency} (note that the dots have been connected by straight lines to facilitate the visualisation). 

The results in Fig.~\ref{fig:ns3-conservativeness} show a positive non-linear relationship between the CI of an algorithm and the energy efficiency achieved for all the devices considered. MP is the algorithm with the lowest CI, which is in consonance with its aggressiveness (i.e., frequent jumps between MCS/TXP values, as shown in Figs.~\ref{fig:ns3-rate} and \ref{fig:ns3-txpower}), and the goodput achieved was also the lowest, as depicted in Fig.~\ref{fig:ns3-goodput}. On the other hand, PARF, PRCS and RRPAA achieved a similar performance in terms of goodput, but the ones with the most conservative behaviour (PRCS and RRPAA, as it can be seen in Figs.~\ref{fig:ns3-rate} and \ref{fig:ns3-txpower}) also achieve both the highest CI and energy efficiency.

This result evidences that the performance gaps uncovered by Fig.~\ref{fig:efficiency-goodput} under optimal conditions have also an impact in real-world RA-TPC algorithms. Therefore, we confirm that this issue must be taken into account in the design of more energy-efficient rate and transmission power control algorithms.

\section{Conclusions}\label{sec:conclusions}

In this paper, we have revisited 802.11 rate adaptation and transmission power control by taking energy consumption into account. While some previous studies pointed out that MIMO rate adaptation is not energy efficient, we have demonstrated through numerical analysis that, even for single spatial streams without interfering traffic, energy consumption and throughput performance are different optimisation objectives. Furthermore, we have validated our results via experimentation.

Our findings show that this trade-off emerges at certain ``mode transitions'' when maximising the goodput, suggesting that small goodput degradations may lead to energy efficiency gains. For instance, a station at the edge of a mode transition may decide to reduce the transmission power a little in order to downgrade the modulation coding scheme. Or an opportunity to achieve a better goodput by increasing the transmission power and modulation coding scheme could be delayed if the expected gain is small.

We have assessed the performance of four state-of-the-art schemes through simulation, and we have demonstrated that certain conservativeness at mode transitions can make a difference for properly designed energy-aware rate adaptation with transmission power control algorithms.

\section*{Acknowledgements}

This work has been performed in the framework of the H2020-ICT-2014-2 projects 5GNORMA (grant agreement no. 671584) and Flex5Gware (grant agreement no. 671563). The authors would like to acknowledge the contributions of their colleagues. This information reflects the consortium's view, but the consortium is not liable for any use that may be made of any of the information contained therein.

\section*{\refname}
\balance
\bibliographystyle{elsarticle-num}
\bibliography{mswim053-ucar}
\end{document}